\documentclass[fleqn,10pt]{article}
\usepackage[utf8]{inputenc}
\usepackage[T1]{fontenc}
\usepackage{bm}
\usepackage{lineno}
\usepackage[superscript,biblabel]{cite}

\RequirePackage[utf8]{inputenc}
\RequirePackage[english]{babel}
\RequirePackage{ifthen}
\RequirePackage{calc}
\AtEndOfClass{\RequirePackage{microtype}}
\usepackage{titling}
\RequirePackage{times}
\RequirePackage{mathptmx} 
\RequirePackage{ifpdf}
\RequirePackage{amsmath,amsfonts,amssymb}
\RequirePackage{graphicx,xcolor}
\RequirePackage{booktabs}
\RequirePackage{authblk}
\RequirePackage[left=2cm,
                right=2cm,
                top=2.25cm,
                bottom=2.25cm,
                headheight=12pt,
                letterpaper]{geometry}                
\RequirePackage[labelfont={bf,sf},
                labelsep=period,
                justification=raggedright]{caption}
\RequirePackage[colorlinks=true, allcolors=blue]{hyperref}
\date{}

\title{Standardization of chemically selective atomic force microscopy for metal-oxide surfaces}

\author[1,2]{Philipp Wiesener}
\author[3]{Stefan Förster}
\author[1,2]{Milena Merkel}
\author[1,2]{Bertram Schulze Lammers}
\author[1,2]{Harald Fuchs}
\author[2,4,5]{Saeed Amirjalayer}
\author[1,2*]{Harry Mönig}
\affil[1]{Universität Münster, Physikalisches Institut, Münster, Germany}
\affil[2]{Center for Nanotechnology, Münster, Germany}
\affil[3]{Martin-Luther-Universität Halle-Wittenberg, Institut für Physik, Halle, Germany}
\affil[4]{Universität Münster, Institut für Festkörpertheorie, Münster, Germany}
\affil[5]{Center for Multiscale Theory and Computation, Münster, Germany}

\affil[*]{e-mail: harry.moenig@uni-muenster.de}

\begin{document}

\maketitle
\thispagestyle{empty}

\section*{ABSTRACT}
\textbf{The complex atomic structures and defects of metal-oxide surfaces are vital for a variety of applications in material science and chemistry. While scanning probe microscopy allows accessing atomic-scale structures in real space, elemental discrimination and defect characterization usually rely on indirect assumptions and extensive theoretical modelling. By investigating a variety of different sample systems with increasing structural complexity and inherent defects, we demonstrate that noncontact atomic force microscopy with an O-terminated copper tip allows imaging metal-oxide surfaces with a clear elemental contrast. This universal approach provides not only immediate access to the metal- and oxygen sub-lattices, but also to the chemical and structural configuration of atomic-scale defect structures. The observed contrast can be explained by purely electrostatic interactions between the negatively charged tip apex and the strongly varying electrostatic potential between metal- and oxygen surface sites. These results offer a standardized methodology for the direct structural characterization of even most-complex metal-oxide surfaces, which is highly relevant for a fundamental understanding of atomic-scale processes on these material systems.}

\section*{Introduction}
The local atomic structure of metal-oxide surfaces largely determines their properties, which are important for a variety of applications ranging from energy conversion and electronics,\cite{Butler2019,Agnoli2018,Coll2019} to data storage,\cite{Lee2009} heterogeneous catalysis,\cite{Garzon2019,Raizada2021} and biomedical sciences.\cite{Limo2018} Enabling real-space investigations on the atomic-level, scanning probe microscopy methods have become key tools in modern material characterization.\cite{Bian2021,Meyer2022} In particular, scanning tunneling microscopy (STM) and noncontact atomic force microscopy (nc-AFM) have provided groundbreaking results with regard to surface processes on metal oxides.\cite{Freund2008,Barth2011,Setvin2017,Parkinson2016,Wagner2021,Hulva2021,Freund2022} Yet, the surface structure determination of such compounds remains a considerable obstacle. Even for relatively simple bulk structures, the heterogeneity can lead to complex surface reconstructions and a high degree of disorder, which makes structural elucidation extremely challenging or even intractable.\cite{Lindsay2022} At the same time, the elemental discrimination based on contrast analysis of scanning probe microscopy images can be extremely difficult. Especially the often unknown identity of the probe-tip decisively determines the image contrast and limits the reproducibility of such experiments.\cite{Monig2013,Sang2014,Schulz2018,Liebig2020,SchuLa2021} As a consequence, elemental recognition on the atomic-scale usually relies on indirect structural considerations, which are prone to errors and require extensive theoretical modelling of experimental contrasts. This in turn requires assumptions not only about the surfaces themselves, but also about the nature of the probe tip.  \cite{Hofer2003,Sang2014,Monig2013,Tautz2019,Wagner2021}

Strongly related to the topography and chemical forces between the tip and the surface, nc-AFM has shown to be a promising tool towards imaging with chemical selectivity. In this regard, a major step was the work by Sugimoto et al. in 2007 for a Si-Pb-Sn alloy model surface, which was achieved by recording force spectra with different tip terminations combined with density functional theory (DFT) simulations.\cite{Sugimoto2007} With a related approach using various Si-terminated tips, Onoda et al. developed a methodology to evaluate the electronegativity of different surface species, embedded in a Si(111) surface, in relation to the probe tip. \cite{Onoda2017} A major breakthrough in nc-AFM was the work by Gross et al., \cite{Gross2009} which initiated atomically defined tip functionalization where inert probe particles (predominantly CO) are attached to a metallic tip. \cite{Mohn2013,Giessibl2021} Using such tip functionalization in conjunction with the small oscillation amplitudes ($\leq 1  $ \AA) of a qPlus force sensor\cite{Giessibl2019} allows imaging organic- and inorganic nano structures with a drastically increased resolution.\cite{Jelinek2017,Giessibl2021,Gross2018} With regard to elemental recognition with CO-tips, Schulz et al. used a combination of nc-AFM, Kelvin-Probe AFM, and DFT-based simulations to understand the contrast on hexagonal boron nitride. \cite{Schulz2018} Liebig et al. demonstrated a site-selective contrast of CO-tips on an ionic CaF$_2$(111) surface, which is based on the negative charge density at the CO-tip apex.\cite{Liebig2020} Van der Heijden et al. employed 3D force spectroscopy to study chemical differences, atomic size- as well as adsorption height effects in organic nano structures.\cite{VanDerHej2016} 

While tip-functionalization by probe particles in nc-AFM has shown groundbreaking success, their weak coupling to the metallic tip base can cause tip flexibility issues and related imaging artefacts.\cite{Gross2012,Pavlicek2012,Hamalainen2014,Hapala2014,Liebig2020} Previously, we established a complementary approach of nc-AFM tip functionalization using an oxygen-terminated copper tip (CuOx-tip). Due to the covalent tetrahedral bonding configuration of the terminal oxygen, CuOx-tips are highly rigid allowing to significantly reduce artefacts.\cite{Monig2018a} A systematic comparison of various atomically defined tip terminations (pure metal-, CO-, Xe-, and CuOx-tips) on a partially oxidized Cu(110) (2x1)O surface revealed not only the absence of tip flexibility artefacts, but also a clear chemical selective (Cu/O)-contrast only for the CuOx-tips.\cite{SchuLa2021} This remarkable result on a metal-oxide surface must be set into context to the long-standing dream in scanning probe microscopy to achieve direct chemical imaging on the single atom level on a free surface.\cite{Jung1995,Shluger1999,Sugimoto2007,Schulz2018} 

In the present work we demonstrate the outstanding ability of CuOx-tips for chemical imaging of metal-oxide surfaces by investigating copper-, silver-, iron-, and titanium-oxide systems with a step-wise increasing structural complexity. For all the specific surfaces, metal- and oxygen sites and sub-lattices can be clearly identified in the constant height nc-AFM images and force spectroscopy data. Comparing well-known surface structures with their calculated local electrostatic potential reveals that the chemically selective contrast is purely based on electrostatic tip-sample interactions. Furthermore, we focus on defect structures and surfaces where so far no conclusive models exist. These results demonstrate that the chemical-selective contrast between metal and oxygen sites is universal and can be generalized for metal-oxide surfaces.

\section*{Results and Discussion}

\begin{figure}[h!t]
\centering
\includegraphics[width=\linewidth]{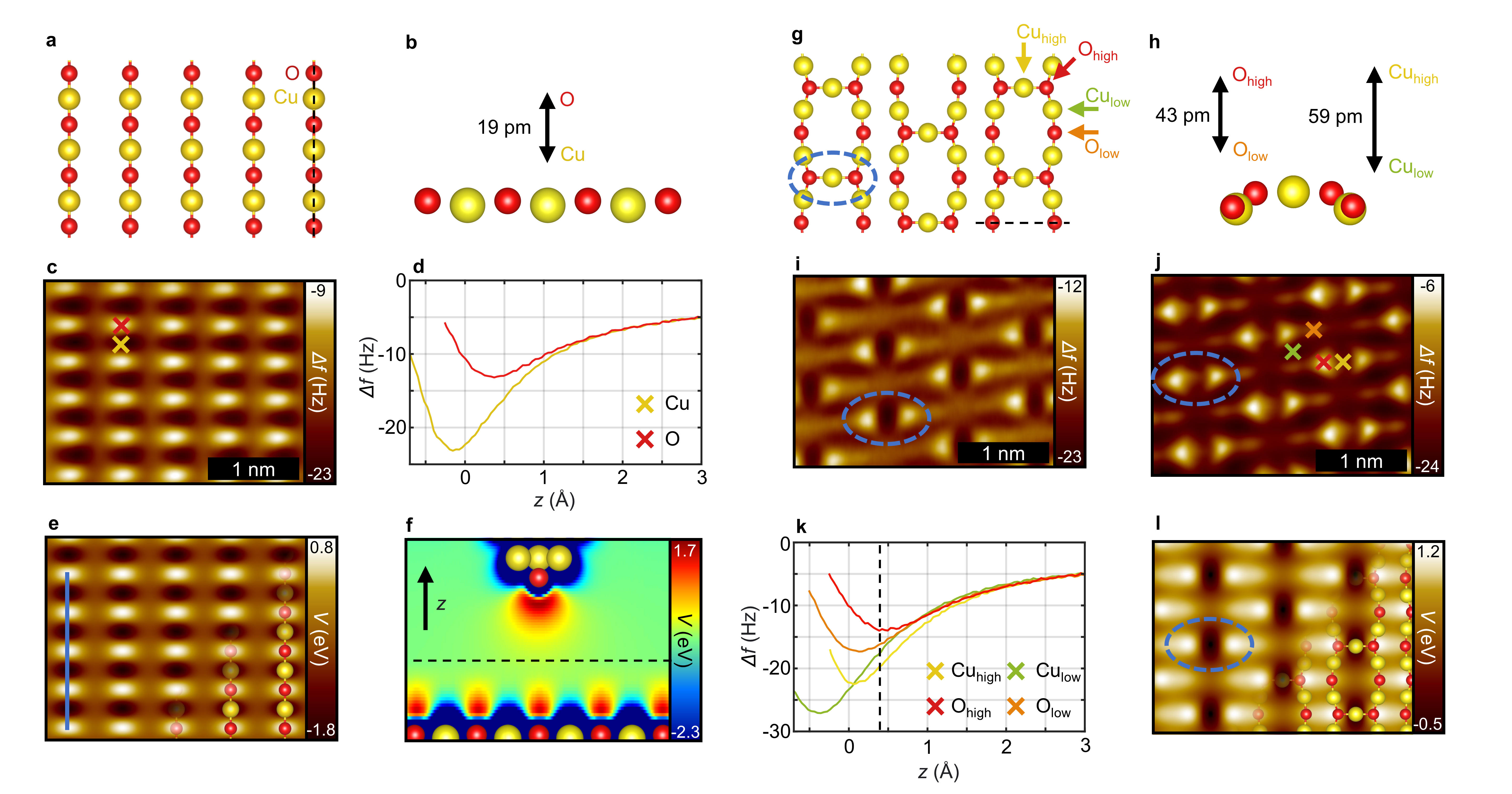}
\caption{\textbf{(2x1)O-reconstruction and (6x2)O-reconstruction on Cu(110).} \textbf{a} | DFT optimized structure of the (2x1)O-reconstruction as top view. For clarity only the top atomic layer is shown. \textbf{b} | Side view along the dashed line in a, depicting the relative heights of the surface species. \textbf{c} | CuOx-tip constant height nc-AFM of the (2x1)O-reconstruction. \textbf{d} | $\Delta f(z)$-spectra of the Cu- and O-sites marked in c. $z=0$ corresponds to the tip height of the measurement in c. \textbf{e} | Calculated electrostatic potential of the (2x1)O-reconstruction, plotted at a height of 81 pm relative to the top most surface atom. \textbf{f} | Cross section of the calculated electrostatic potential of the CuOx-tip and the (2x1)O surface. As indicated by the dashed line, the data for the tip and the surface are obtained from separate DFT calculations. \textbf{g} | DFT-optimized structure of the (6x2)O-reconstruction as top view. \textbf{h} | Side view along the dashed line in g, depicting the relative heights of the surface species. \textbf{i} | CuOx-tip constant height nc-AFM of the (6x2)O-reconstruction (\textbf{j} with reduced tip height). \textbf{k} | $\Delta f(z)$-spectra of the Cu- and O-sites marked in j. $z=0$ corresponds to the tip height of the measurement in j, while the dashed line marks the tip height of the measurement in i. \textbf{l} | Calculated electrostatic potential of the (6x2)O-reconstruction, plotted at a height of 81 pm relative to the top most surface atom.}
\label{Cu110}
\end{figure}

One of the most basic metal-oxide surfaces is the (2x1)O-reconstruction on Cu(110), which consists of rows of alternating copper- and oxygen atoms as depicted by a DFT-optimized structure in Fig. \ref{Cu110}a and b. A typical CuOx-tip nc-AFM measurement recorded in constant height mode is shown in Fig. \ref{Cu110}c. It clearly shows the O atoms as bright (repulsive) protrusions and the Cu atoms as dark (attractive) depressions.\cite{SchuLa2021} This chemical selectivity is further emphasized by $\Delta f(z)$-spectra recorded on the O- and Cu sites (Fig. \ref{Cu110}d), where a distinct separation of the curves over an extended height range can be observed. Specifically, the strongest chemical contrast is obtained at a tip distance $z$ around the $\Delta f(z)$-minimum on the Cu-site. To further understand the contrast mechanism, Fig. \ref{Cu110}e shows the calculated electrostatic potential of the surface based on the optimized structure. The remarkable agreement with the experimentally obtained contrast (Fig. \ref{Cu110}c) indicates that the chemical selectivity mainly relies on electrostatic interactions between the negative charge of the tip terminating O-atom and the strongly varying charge density of metal- and oxygen sites on the surface. As visualized by a cross section of the electrostatic potential within the tip-sample junction (Fig. \ref{Cu110}f): Metal atoms are dominated by electrostatic attraction, while the oxygen atoms show a pronounced repulsive interaction with the CuOx-tip. 

For a step-wise increase of the structural complexity we investigate the well-known (6x2)O-reconstruction on Cu(110).\cite{Duan2010} It features various metal- and oxygen atoms at different relative heights (Fig. \ref{Cu110}g and h) and allows assessing how the chemical selective contrast is affected by topography. Similar to the (2x1)O-reconstruction, the (6x2)O-surface comprises rows of alternating copper- (Cu$_{\text{low}}$) and oxygen atoms (O$_{\text{low}}$). In contrast, however, the higher degree of oxidation leads to additional topographically higher copper atoms (Cu$_{\text{high}}$) located in between these rows, elevating the neighboring oxygen atoms (O$_{\text{high}}$).\cite{Duan2010} The resulting height differences are depicted in the side-view of the DFT-optimized structure in Fig. \ref{Cu110}h. At larger tip-sample distances, the nc-AFM measurement (Fig. \ref{Cu110}i) is dominated by the topographically elevated Cu$_{\text{high}}$- and O$_{\text{high}}$ sites (blue circle). Again, a site selective contrast arises for the attractive Cu$_{\text{high}}$ atoms and their neighboring, repulsive O$_{\text{high}}$ atoms. By approaching further to the surface (Fig. \ref{Cu110}j), a similar chemical selective contrast emerges for the underlying rows of Cu$_{\text{low}}$ and O$_{\text{low}}$ atoms, while the elevated Cu$_{\text{high}}$ atoms appear more repulsive due to increasing Pauli repulsion. The corresponding $\Delta f(z)$-spectra of each surface species in Fig. \ref{Cu110}k highlight the height-dependent evolution of the contrast. Despite the distinct height differences, a clear separation of the $\Delta f(z)$-spectra for metal and oxygen sites is revealed. Importantly, even the topographically higher Cu$_{\text{high}}$ atom remains distinctly more attractive than the topographically lower oxygen sites, indicating a robust height sensitivity of the chemical-selective contrast. Figure \ref{Cu110}l shows the electrostatic potential for the (6x2)O-reconstruction, which is dominated by the topographically higher Cu$_{\text{high}}$ atoms. Due to their reduced electron density, they appear as dark spots whereas both oxygen sites appear bright due to an increased electron density, which agrees well with the nc-AFM contrast. In the following, the apparent close relation between CuOx-tip nc-AFM measurements and the electrostatic potential is further investigated by focusing on the (2$\sqrt{2}$ x $\sqrt{2}$)R45$^{\circ}$O- and c(2x2)N-reconstructions on Cu(100).

\begin{figure}[h!t]
\centering
\includegraphics[width=0.5\linewidth]{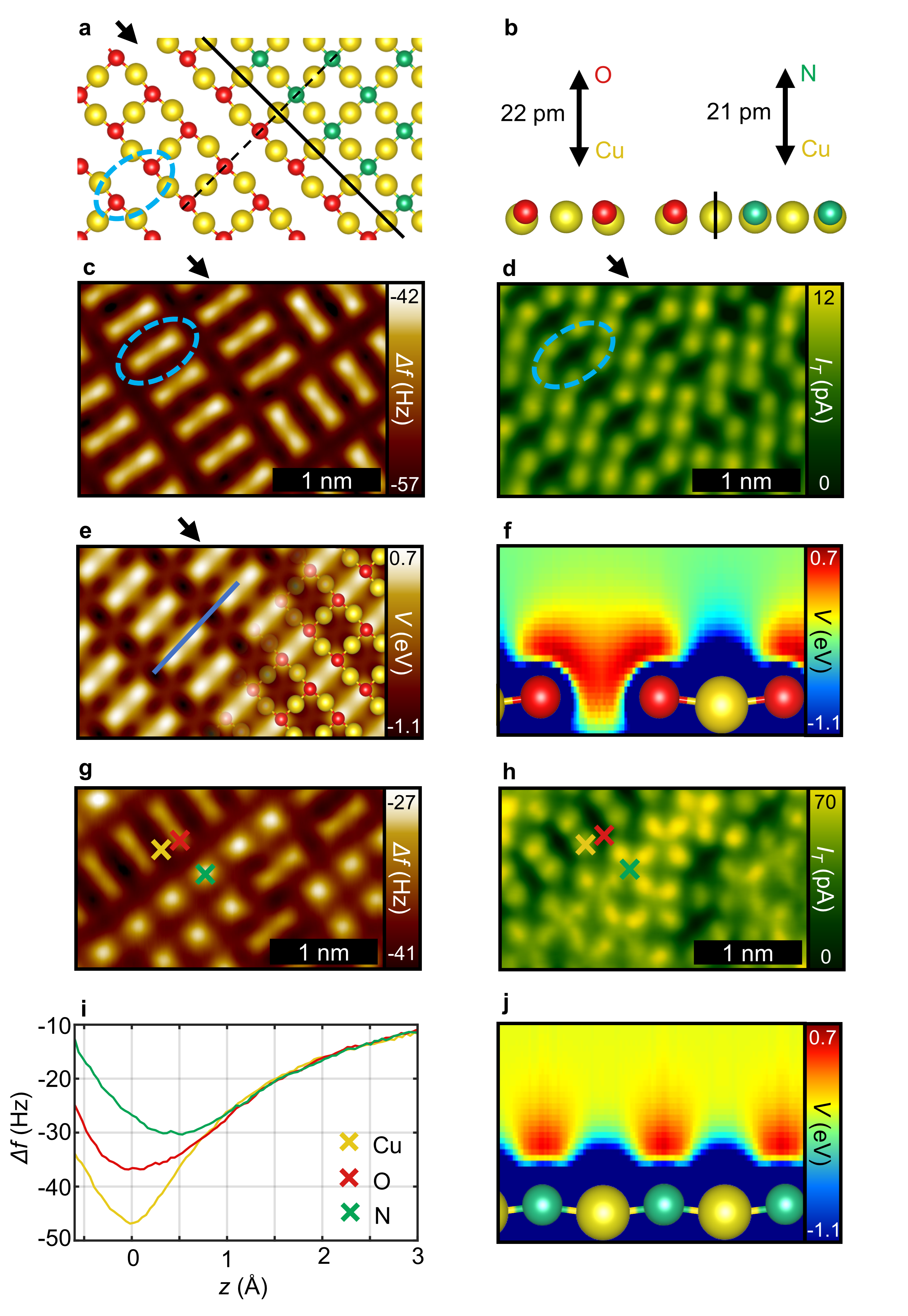}
\caption{\textbf{(2$\sqrt{\textbf{2}}$ x $\sqrt{\textbf{2}}$)R45$^{\circ}$O-reconstruction and c(2x2)N-reconstruction on Cu(100).} \textbf{a} | DFT-optimized structure of the (2$\sqrt{2}$ x $\sqrt{2}$)R45$^{\circ}$O-reconstruction (bottom left) and c(2x2)N-reconstruction (top right) as top view. For clarity only the top atomic layer is shown. As indicated by the black line, the structure for the copper-oxide and the copper-nitride systems are obtained from separate DFT calculations. \textbf{b} Side view along the dashed line in a, depicting the relative heights of the surface species. \textbf{c} | CuOx-tip constant height nc-AFM measurement of the (2$\sqrt{2}$ x $\sqrt{2}$)R45$^{\circ}$O-reconstruction. \textbf{d} | Simultaneously recorded tunneling current. \textbf{e} | Calculated electrostatic potential plotted at a height of 82 pm relative to the top most surface atom. \textbf{f} | Cross section of the electrostatic potential along the blue line shown in e. \textbf{g} | CuOx-tip constant height nc-AFM measurement of the co-deposited (2$\sqrt{2}$ x $\sqrt{2}$)R45$^{\circ}$O-reconstruction and c(2x2)N-reconstruction. \textbf{h} | Simultaneously recorded tunneling current. \textbf{i} | $\Delta f(z)$-spectra of the Cu-, O- and N-sites marked in g. $z=0$ corresponds to the tip height of the measurement in g. \textbf{f} | Cross section of the electrostatic potential for the c(2x2)N-reconstruction along the same axis as in f.}
\label{Cu100}
\end{figure}

The Cu(100) (2$\sqrt{2}$ x $\sqrt{2}$)R45$^{\circ}$O surface system again features rows of copper and oxygen atoms, which, however, are separated by alternating added (AR) and missing (MR) rows of copper atoms (Fig. \ref{Cu100}a lower left and Fig. \ref{Cu100}b left). Contrary to the previous copper oxide systems, this surface shows two domain orientations with identical atomic structure along the [010] and [001] directions, respectively.\cite{Monig2013}
In the nc-AFM measurement in Fig. \ref{Cu100}c elongated protrusions of the oxygen sites emerge (blue circles in Fig. \ref{Cu100}), which bridge the missing rows (black arrows in Fig. \ref{Cu100}), whereas the topographically lower Cu sites appear as weak depressions surrounding the oxygen protrusions. The location of these Cu atoms is further confirmed by an increased tunneling current in the simultaneously recorded STM data (Fig. \ref{Cu100}d). Again, the calculated electrostatic potential in Fig. \ref{Cu100}e shows very good agreement with contrast signatures observed in the nc-AFM measurement (Fig. \ref{Cu100}c), which holds especially for the dimer-like shape of the oxygen atoms. Further examining a vertical cross section of the electrostatic potential along the dimer axis (Fig. \ref{Cu100}f) illustrates how the intermediate missing row distorts the valence electron orbitals of the oxygen atoms leading to the dimer-like charge density.

As shown by Wofford et al.,\cite{Wofford} the (2$\sqrt{2}$ x $\sqrt{2}$)R45$^{\circ}$O-structure can coexist with the nitrogen-induced c(2x2)N-reconstruction\cite{Giessibl14,Choi08} on the Cu(100) surface. In the latter case, the nitrogen atoms occupy hollow sites on Cu(100) without the formation of missing or added rows (Fig. \ref{Cu100}a upper right, Fig. \ref{Cu100}b right, Supplementary Fig. S1). Figure \ref{Cu100}g displays an nc-AFM image encompassing both phases, where the dimer-like protrusions of the oxygen atoms coexist with isolated maxima of the nitrogen atoms. Additionally, the simultaneously recorded tunneling current in Fig. \ref{Cu100}h allows to clearly allocate the metal sub-lattice for both reconstructions (please note the structure of the phase boundary in Fig. \ref{Cu100}a and b as deduced from these measurements). Corresponding $\Delta f(z)$-spectra are depicted in Fig. \ref{Cu100}i, which show a clear separation even between the N- and O-spectra. Again, these findings excellently correlate with the cross sections of the calculated electrostatic potential. In fact, in case of the O-dimers, the charge density laterally merges from both sides towards the missing row (Fig. \ref{Cu100}f), while the charge density above the N-sites vertically spreads from the surface (Fig. \ref{Cu100}j). Importantly, the results discussed so far demonstrate that the nc-AFM contrast with CuOx-tips strongly correlates with the local electrostatic potential and the inherent charge density contours, which is the basis for the observed chemically selective contrast.

Contrary to the well-known surfaces discussed above, an approach of the structural elucidation of two coexisting O-induced reconstructions on Ag(111) is given in the Supplementary Information. Their atomic structures are largely unknown, thus they offer an opportunity to explore the applicability of the described chemical selectivity. One measurement addresses the p(4x4)O-reconstruction, where several contradicting models based on combined STM/DFT data are discussed in literature. \cite{Andry2024,AgO} Our CuOx-tip nc-AFM measurement shown in Supplementary Fig. S2a-d allows developing a completely new model based on AgO$_4$ units with different O-orientations with respect to the substrate. The second surface structure we investigated, shows a striped STM contrast along the Ag(111) high-symmetry axes. While previous studies assumed that such stripes are built from protruding Ag atoms, \cite{AgO} our measurement strongly suggests the absence of metal atoms within the top atomic layer (Supplementary Fig. S2e-h). Rather, the CuOx-tip nc-AFM contrast features a pore arrangement of protruding oxygen atoms, while the simultaneously recorded tunnelling current reveals the underlying Ag sub-lattice featuring a vacancy structure based on the Ag(111) unit cell.

\begin{figure}[h!t]
\centering
\includegraphics[width=\linewidth]{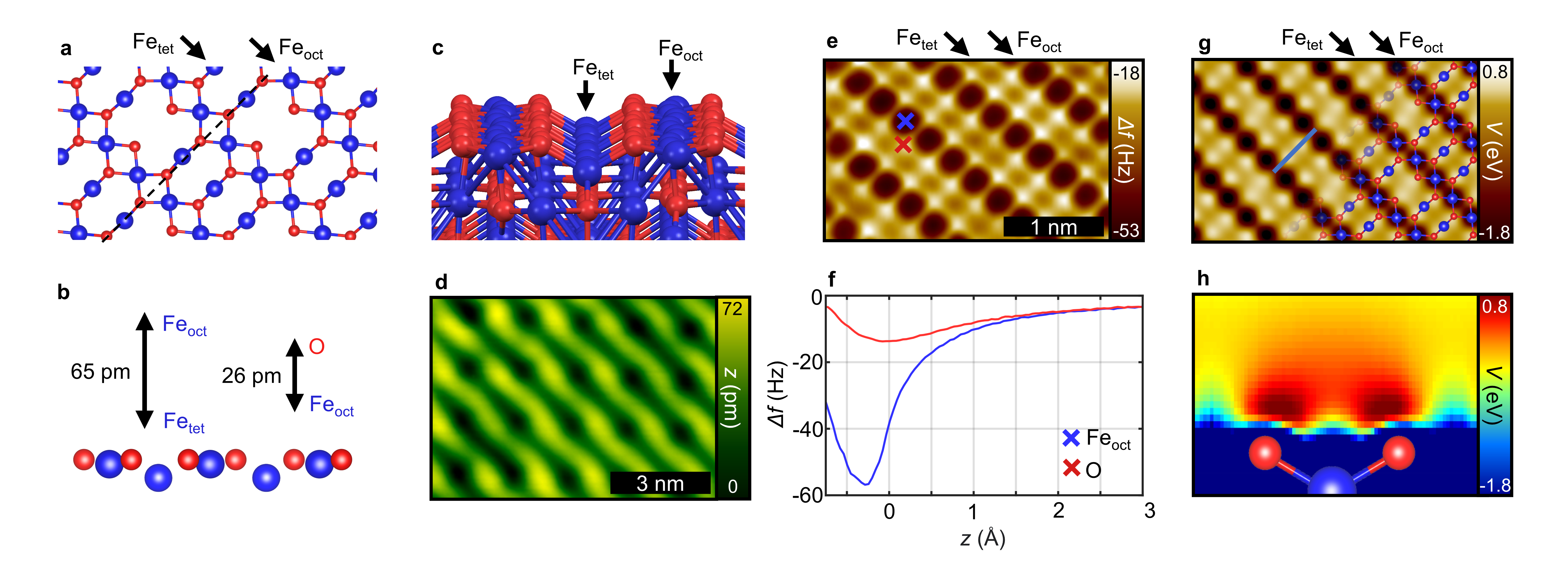}
\caption{\textbf{Magnetite $\textbf{Fe}_{3}\textbf{O}_{4}$(001).} \textbf{a} | DFT-optimized structure of the subsurface cation vacancy model by Bliem et al.\protect\cite{Bliem2014} as top view. For clarity only the top atomic layer is shown. \textbf{b} | Side view along the dashed line in a, depicting the relative heights of the surface species. \textbf{c} | Perspective of the DFT-optimized structure, which additionally shows the subsurface layer. \textbf{d} | STM Overview image with Feedback-loop (1.0 V, 10 pA). \textbf{e} | CuOx-tip constant height nc-AFM. \textbf{f} | $\Delta f(z)$-spectra of the Fe- and O-sites marked in e. $z=0$ corresponds to the tip height of the measurement in e. \textbf{g} | Calculated electrostatic potential plotted at a height of 83 pm relative to the top most surface atom. \textbf{h} | Cross section of the electrostatic potential along the blue line in g.}
\label{Mag100}
\end{figure}

To generalize our approach to further elements, two different iron-oxide surfaces are considered. The magnetite $\text{Fe}_{3}\text{O}_{4}$(001) surface is already well-described by the subsurface cation vacancy model \cite{Bliem2014} and holds relevance in various catalytic applications.\cite{Parkinson2016} The DFT-optimized structure is shown in Fig. \ref{Mag100}a-c, consisting of alternating rows of octahedral (Fe$_{\text{oct}}$) and tetrahedral iron atoms (Fe$_{\text{tet}}$). While the Fe$_{\text{oct}}$ are almost co planar with the O atoms, the Fe$_{\text{tet}}$ are situated 65 pm lower (Fig. \ref{Mag100}b). In the suggested model, subsurface Fe-vacancies reorganize the second layered row of Fe$_{\text{tet}}$, resulting in an unoccupied tetrahedral position every fourth atom. This establishes the ($\sqrt{2}$ x $\sqrt{2}$)R45$^{\circ}$ periodicity and results in the characteristic distorted STM contrast depicted in Fig. \ref{Mag100}d. The nc-AFM measurement with a CuOx-tip in Fig. \ref{Mag100}e reveals a distinct contrast between the attractive Fe$_{\text{oct}}$ atoms (dark spots) and their surrounding four O atoms imaged as repulsive maxima (bright spots). The $\Delta f(z)$-spectra in Fig. \ref{Mag100}f feature the typical separation of the curves demonstrating a chemical selectivity over an extended $z$-range, very similar as found for the copper oxide surfaces above. Interestingly, no significant contrast features are observed for the topographically lower Fe$_{\text{tet}}$ atoms. Notably, this finding is also reflected in the calculated electrostatic potential (Fig. \ref{Mag100}g), which again shows excellent agreement with the measurement. An examination of a cross section of the electrostatic potential along the O-Fe$_{\rm{tet}}$-O axis (Fig. \ref{Mag100}h) further reveals how the valence electron orbitals and related charge density of the oxygen atoms shield the topographically lower Fe$_{\rm{tet}}$ atoms towards the surface normal, whereby imaging of these atoms is restricted.

The second iron-oxide system we investigated concerns a surface reconstruction of $\text{Fe}_{3}\text{O}_{4}$(110), where the structure is largely unknown. Under standard preparation conditions $\text{Fe}_{3}\text{O}_{4}$(110) forms [111]-oriented nano-facets, which leads to a strongly corrugated topography in STM experiments. \cite{Parkinson2016} However, for higher annealing temperatures, this structure can coexists with an atomically flat reconstructed phase.\cite{jansen1995,Walls2016} Although constant-height nc-AFM imaging was challenging due to mobile species on this phase, its basic structure could be revealed (Supplementary Fig. S3). 

\begin{figure}[h!t]
\centering
\includegraphics[width=0.5\linewidth]{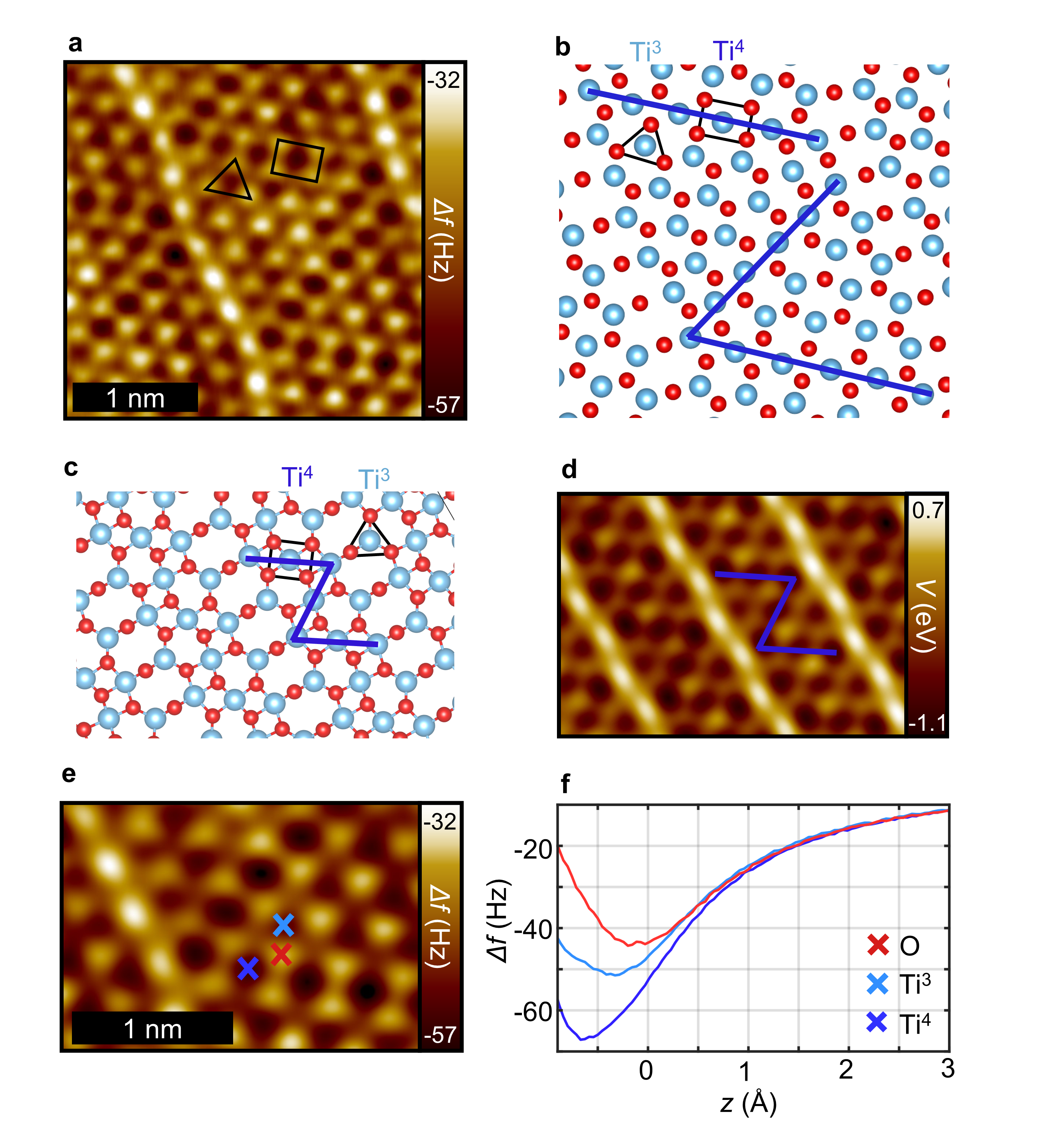}
\caption{\textbf{z-TiO$_{\rm{x}}$-phase on Pt(111).} \textbf{a} | CuOx-tip constant height nc-AFM of the zigzag structure. \textbf{b} | Extracted atomic positions from an area of the measurement. The blue line marks the zigzag-like arrangement of the fourfold coordinated Ti$^{\rm{4}}$ atoms (square). The threefold coordinated Ti$^{\rm{3}}$ atoms are marked with a triangle. \textbf{c} | DFT-optimized structure by Barcaro et al.\protect\cite{Barcaro2007} of a similar structure to the measurement but with a smaller unit cell as top view. For clarity only the top atomic layer is shown.  \textbf{d} | Calculated electrostatic potential of the DFT-optimized structure in c, plotted at a height of 83 pm relative to the top most surface atom. \textbf{e} | Excerpt of the CuOx-tip constant height nc-AFM. \textbf{f} | $\Delta f(z)$-spectra of the different surface sites marked in e. $z=0$ corresponds to the tip height of the measurement in a.}
\label{complexsurfaces}
\end{figure}

To apply our methodology to a surface with a much higher structural complexity compared to the systems considered so far, we also investigated a titanium oxide nano-structure. TiO$_{\rm{x}}$ thin films grown on Pt(111) show various stable surface phases \cite{Dulub2000,Sedona2005,Barcaro2009} where we focus on one of the so called zigzag (z-) phase. Thereby, the label "zigzag" refers to a distinct contrast found in STM experiments (see also Supplementary Fig. S4). The z-phase exist with unit cells of different size and stochiometry where various models are discussed in literature.\cite{Jennison2001,Sedona2005,Barcaro2007,Barcaro2009} An nc-AFM measurement of such a zigzag structure in its maximal complexity is shown in Fig. \ref{complexsurfaces}a. Based on the results discussed above, the oxygen- and titanium sub-lattices can readily be identified. The atomic positions directly extracted from the measurement are depicted in Fig. \ref{complexsurfaces}b showing the complex arrangement of threefold- and fourfold coordinated titanium atoms (Ti$^{\rm{3}}$ and Ti$^{\rm{4}}$, respectively) and twofold- and threefold coordinated oxygen atoms. The line segments of the zigzag structure (indicated as blue lines in Fig. \ref{complexsurfaces}b) possess a length of 6 Ti$^{\rm{4}}$ atoms. The areas between these lines are filled by differing numbers of Ti$^{\rm{3}}$ units. Moreover, the periodically repeating zigzag structures are separated by rows of elevated twofold-coordinated oxygen atoms noticed by their strongly repulsive contrast and a slight wave-like appearance. This arrangement is in agreement with the DFT calculation by Barcaro et al. of a similar structure but with a smaller unit cell, i.e. with zigzag line segments of a length of 3 Ti$^{\rm{4}}$ atoms shown in Fig. \ref{complexsurfaces}c.\cite{Barcaro2007} Nevertheless, the corresponding electrostatic potential shown in Fig. \ref{complexsurfaces}d reproduces the most dominant contrast features observed for the more complex structure found in the experiment. Figures \ref{complexsurfaces}e and f show $\Delta f(z)$-spectra recorded above the marked sites in e. Remarkably, the Ti$^{\rm{4}}$ atoms appear more attractive than the Ti$^{\rm{3}}$ species. Yet, according to the DFT structure mentioned above, these atoms are assumed to be nearly in the same height range ($\pm 3$ pm)\cite{Barcaro2007} (Supplementary Fig. S4). Presumably, the more attractive interaction above the Ti$^{\rm{4}}$ sites is not an effect of topography but from the reduced charge density as a consequence of their higher coordination. 

\begin{figure}[h!t]
\centering
\includegraphics[width=\linewidth]{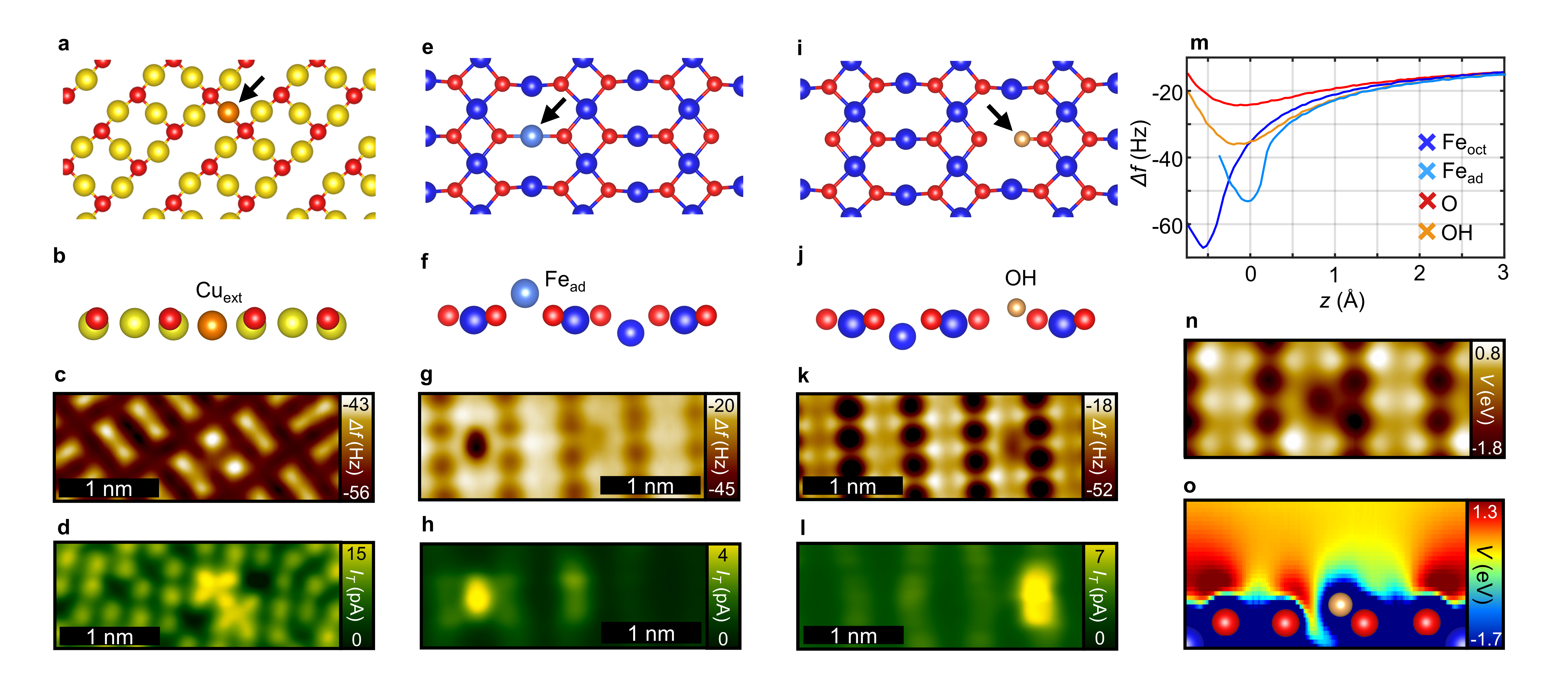}
\caption{\textbf{Defect structures.} \textbf{a} | Structural model of the Cu$_{\rm{ext}}$-defect of the (2$\sqrt{\textbf{2}}$ x $\sqrt{\textbf{2}}$)R45$^{\circ}$O-reconstruction on Cu(100) as top view. | \textbf{b} Side view from a. \textbf{c} | CuOx-tip constant height nc-AFM of the Cu$_{\rm{ext}}$-defect. \textbf{d} | Simultaneously recorded tunneling current. \textbf{e} | Structural model of the Fe$_{\rm{ad}}$-defect on $\text{Fe}_{3}\text{O}_{4}$(001) as top view.  \textbf{f} | Side view from e. \textbf{g} | CuOx-tip constant height nc-AFM of the Fe$_{\rm{ad}}$-defect. \textbf{h} | Simultaneously recorded tunneling current ($U=0.17$ V). \textbf{i} | DFT-optimized structure of the OH-defect on $\text{Fe}_{3}\text{O}_{4}$(001) as top view. \textbf{j} | Side view from i. \textbf{k} | CuOx-tip constant height nc-AFM of the OH-defect. \textbf{l} | Simultaneously recorded tunneling current ($U=0.17$ V). \textbf{m} | $\Delta f(z)$-spectra of the Fe-, O- and defect-sites in g and k. $z=0$ corresponds to the tip height of the measurement in g. \textbf{n} | Calculated electrostatic potential, plotted at a height of 83 pm relative to the top most oxygen atom. \textbf{o} | Cross section of the electrostatic potential along the axis of the OH-group.}
\label{defects}
\end{figure}

Besides the periodic structures of the metal-oxide systems studied so far, the characterization of atomic point defects is of crucial importance for their macroscopic catalytic, optical, and electronic properties.\cite{Pastor2022,Setvin2017,Nilius2015,Raizada2021,DIEBOLD2003,Parkinson2016} In the following we exemplify how the chemical selectivity of CuOx-tips can be used to characterize defects, which we observed in the systems discussed above. One example can be seen within Fig. \ref{defects}a-d, which shows a frequently occurring filled-row defect of the (2$\sqrt{2}$ x $\sqrt{2}$)R45$^{\circ}$O-reconstruction \cite{Baykara2013} (defect free surface in Fig. \ref{Cu100}). In the present case a single extra Cu atom (Cu$_{\rm{ext}}$) fills a missing row site between two oxygen atoms. The related charge reorganization disrupts the typical dimer-like structure of the neighboring oxygen atoms leading to two individual protrusions, which appear considerably brighter compared to the dimer-like oxygen pairs in the nc-AFM measurement (Fig. \ref{defects}c). The presence of the extra Cu$_{\rm{ext}}$ is further confirmed by the simultaneously recorded STM image in Fig. \ref{defects}d, which allows to clearly allocate the metal sub-lattice. 

Furthermore, an iron ad-atom defect (Fe$_{\rm{ad}}$) can be observed on the $\text{Fe}_{3}\text{O}_{4}$(001) surface (Fig. \ref{defects}e-h, defect free surface in Fig. \ref{Mag100}). The metallic nature of this common defect \cite{Bliem2014} can be identified by the strongly attractive tip-sample interaction, and further confirmed by an increased tunneling current (Fig. \ref{defects}h). In  addition, The atomic structure of the underlying $\text{Fe}_{3}\text{O}_{4}$(001) lattice is clearly visible, which allows to confirm that the Fe$_{\rm{ad}}$ occupies the bulk continuation sites of the Fe$_{\rm{tet}}$ atoms. \cite{Bliem2014}

For the defect chemistry of metal oxide materials, hydrogen, which usually forms hydroxyl species, plays an important role.\cite{Li2021} On the $\text{Fe}_{3}\text{O}_{4}$(001) surface, for instance, hydrogen located at the highly reactive bulk continuation sites of the Fe$_{\rm{tet}}$ atoms (Fig. \ref{defects}i and j) is the most commonly observed defect on freshly prepared samples.  \cite{Parkinson2016} An nc-AFM image of such a defect is seen in Fig. \ref{defects}k with the corresponding $\Delta f(z)$-spectra in Fig. \ref{defects}m. The formed OH-group locally reduces the charge density leading to a significantly more attractive nc-AFM contrast as compared to the O atoms. The simultaneously recorded STM image in Fig. \ref{defects}l further confirms this defect, as the neighboring iron atoms exhibit a higher tunneling current due to more empty states, as reported in previous studies.\cite{Parkinson2016} Again, the nc-AFM contrast is excellently reproduced by the calculated electrostatic potential in Fig. \ref{defects}n and o, based on the corresponding DFT-optimized structure. 

\section*{Conclusions}
By investigating a variety of different metal-oxide surfaces with increasing structural complexity, we demonstrate that nc-AFM imaging with CuOx-functionalized tips allows to directly reveal the metal- and oxygen sub-lattices. Even for systems with atoms at various relative heights as well as surfaces with a high degree of structural complexity or atomic-scale defects, we obtain a direct assignment of the atomic configurations. The excellent agreement between the nc-AFM data and the calculated electrostatic potential allows explaining the contrast mechanism by the negative charge at the O-terminated tip apex \cite{Monig2018a} and its interaction with the charge density of the surfaces. The successful application of the methodology to defects and surface systems where so far no conclusive models exist, demonstrates the universal applicability of the experimental approach. Our results not only establish CuOx-tips for the chemically selective characterization of metal-oxide surfaces but also open a pathway for the standardization of chemical surface imaging techniques at the atomic scale. 

\section*{Methods}

A low-temperature scanning probe microscope (SPM) with the MATRIX SPM control system (LT-STM/AFM) from Scienta Omicron was used for the experiments. The microscope was operated under ultra-high vacuum conditions with a base pressure below 5 x 10$^{-11}$ mbar and cooled to 78 K by a liquid nitrogen bath cryostat. The used qPlus force sensors\cite{Giessibl2019} enabled simultaneous STM and AFM data recording with resonance frequencies f$_{0}$ of 26-28 kHz and quality factors of 5k–15k. Unless otherwise stated, the AFM experiments were performed with a constant amplitude (1.0 \AA) active feedback loop and in constant height mode with a bias voltage of 0 V. Raw image data were processed with a Gaussian filter (Scanning Probe Image Processor, SPIP™ 5.1).

To form a CuOx-tip, chemically and focused ion beam (FIB) etched tungsten tips with a tip diameter of less than 30 nm were used. By voltage pulses up to 6 V and indentations up to 2 nm on an Au(111) surface, a clean metallic tip apex is shaped. Afterwards, the tip forming process was continued on the partially oxidized Cu(110) (2x1)O-surface (preparation see below). With a set point of 100 mV and 50 pA in STM feedback, a copper-oxide cluster is picked up by voltage pulses up to 4 V and indentations up to 4 nm into the oxide stripes. By further indentations within a range of 0.2 to 0.6 nm, the copper oxide cluster is then formed into a CuOx-tip. Before recording all of the shown data, a constant height nc-AFM measurement of an oxide stripe on the Cu(110) (2x1)O-surface according to Fig. \ref{Cu110}c is taken to confirm the covalent tetrahedral bonding configuration of the terminal oxygen and its typical imaging contrast.\cite{Monig2018a,SchuLa2021}

For the sample preparation, the Cu(110), Cu(100), Fe$_{3}$O$_{4}$(001) and Pt(111) single crystals were cleaned by cycles of Ar+ sputtering and subsequent annealing up to 600 K, each for 10 min, before the oxidation processes were performed. For the specific surface reconstructions, a detailed instruction is given below. 

\textbf{Cu(110)-reconstructions}. After the final annealing step of the cleaning procedure the sample temperature was lowered to 450 K. At this temperature molecular O$_{2}$ was dosed with a pressure of 2 x 10$^{-8}$ mbar for 10 to 15 s to induce the (2x1)O-reconstructed oxide stripes. For the (6x2)O-reconstruction, which requires further oxidation, the cleaned Cu(110) crystal was heated to 500 K, before molecular O$_{2}$ was dosed with a pressure of 5 x 10$^{-6}$ mbar for 10 min.

\textbf{Cu(100)-reconstructions}. After cleaning, the Cu(100) crystal was heated to 400 K and molecular O$_{2}$ was dosed with a pressure of 1 x 10$^{-6}$ mbar for 20 min to achieve the (2$\sqrt{2}$ x $\sqrt{2}$)R45$^{\circ}$O-reconstruction. For the c(2x2)N-reconstruction, the cleaned Cu(100) crystal was sputtered with nitrogen at 500 eV for 30 s, followed by annealing at 500 K for 10 min. For the co-deposited surface measured in Fig. \ref{Cu100}g, subsequently, molecular O$_{2}$ was dosed with a pressure of 1 x 10$^{-6}$ mbar for 5 min at an annealing temperature of 400 K.

\textbf{Fe$_{3}$O$_{4}$(001)}. Ar+ sputtering and UHV-annealing might reduce the O/Fe ratio which hinders the surface to reconstruct. To prevent this, for the last cleaning cycle the annealing temperature was increased to 900 K for 30 min. In the first 20 min of annealing, molecular oxygen was dosed with 5 x 10$^{-7}$ mbar, while for the last 10 min, UHV-annealing was performed.

\textbf{TiO$_{\rm{x}}$ thin films}. Titanium was deposited on the clean Pt(111) single crystal by electron-beam evaporation of a Ti rod with a deposition rate of 1 ML per minute for 40-50 s. Afterwards O$_{2}$ was dosed at a sample temperature of 900 K and a pressure of 1 x 10$^{-5}$ mbar for 10 min to form TiO$_{\rm{x}}$ structures on the surface. For the measured zigzag phase in Fig. \ref{complexsurfaces}, the surface must be reduced by UHV annealing at 1050 K for 10 min.

\textbf{Computational details}. Utilizing the Vienna Ab Initio Simulation Package (VASP) for periodic density functional theory (DFT) calculations,\cite{VASP1,VASP2} atomistic investigations were conducted on the metal-oxide surfaces focusing on the local potential $V (r)$ calculated as
\[V (r) = V_{\rm{ionic}} (r) + \int \frac{n(r')}{|r-r' |}dr'\]
with the electron density $n(r)$, the ionic potential $V_{\rm{ionic}} (r)$ and the Hartree potential (second term). In all calculations, the GGA function Perdew-Burke-Ernzerhof (PBE),\cite{PBE} together with projected augmented wave (PAW) pseudo-potentials and a plane-wave cutoff of 600 eV for the wave functions, was employed. A Fermi smearing with a standard deviation of 0.2 eV described the electronic states. Dispersion effects were taken into account by using the DFT-D3 method with Becke-Johnson damping function.\cite{Becke-Johnson} Monkhorst Pack k-point sampling with an 8x8x1 mesh was employed for Brillouin zone integration in all calculations. For the copper-based systems, including the CuOx-tip, the geometries were fully optimized (keeping only the bottom layers at the bulk lattice constant of 3.64 \AA). Except for the OH-defect on Fe$_{3}$O$_{4}$(001), which was optimized, the geometry for the Fe$_{3}$O$_{4}$(001) and TiO$_{\rm{x}}$ surfaces were taken from Bliem et al.\cite{Bliem2014} and Barcaro et al.,\cite{Barcaro2007} respectively. For all calculations, a dipole correction to the energy for the slab model was carried out perpendicular to the surface. The convergence criteria for the forces on the nuclei were 0.01 eV \AA$^{-1}$ for the geometry optimization and $10^{-8}$ eV for the electronic relaxation in all calculations. \\

\noindent\textbf{Acknowledgements}\\
This work was funded by the Deutsche Forschungsgemeinschaft (DFG, German Research Foundation) through project 527214857. The authors acknowledge supportive discussions with Martin Setvin.\\

\noindent\textbf{Author contributions}\\
H.M., H.F., and S.A. conceived the project. P.W., S.F., M.M., and B.SL. performed the sample preparations. P.W. conducted the nc-AFM measurements and S.A. carried out the DFT simulations. All authors contributed to data interpretation and manuscript writing drafted by P.W. and H.M..\\

\noindent\textbf{Competing interests}\\
The authors declare no competing interests. \\

\bibliographystyle{unsrt}
\bibliography{sample} 

\newpage
\appendix

\title{Supplementary Information: Standardization of chemically selective atomic force microscopy for metal-oxide surfaces}
\maketitle
\renewcommand{\figurename}{Figure S}
\setcounter{figure}{0}
\setcounter{page}{1}

\begin{figure}[h!]
\centering
\includegraphics[width=0.66\linewidth]{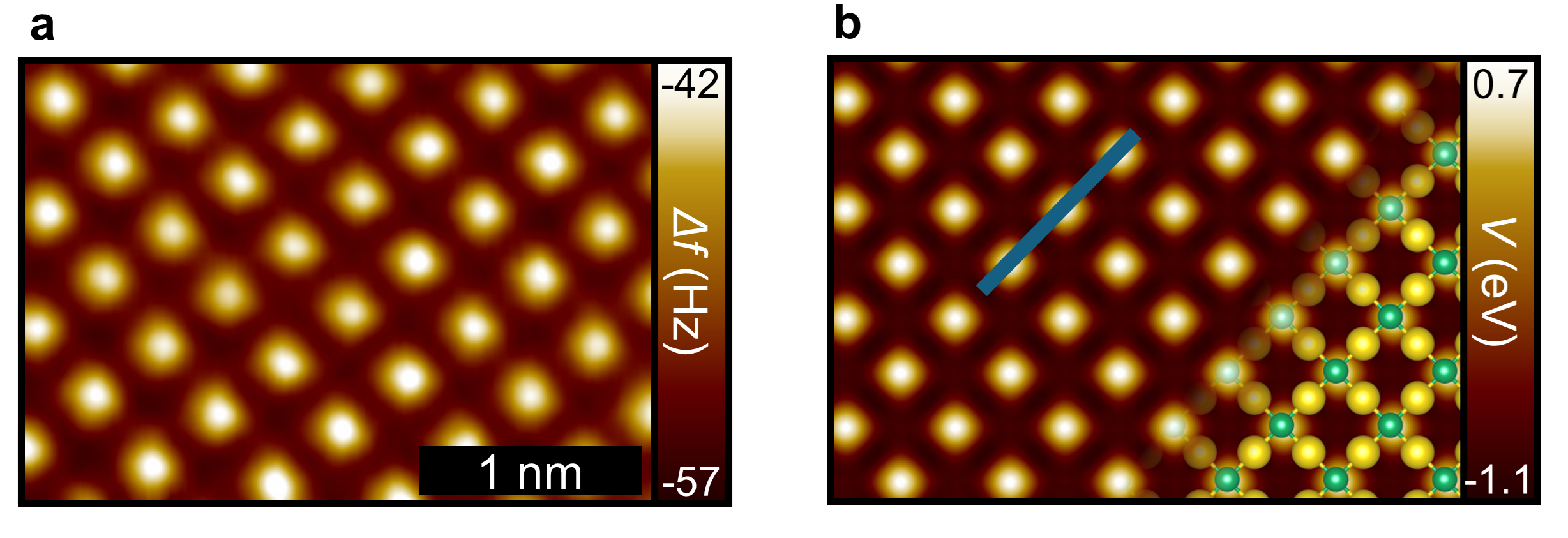}
\caption{\textbf{c(2x2)N-reconstruction on Cu(100).} \textbf{a} | CuOx-tip constant height nc-AFM of the c(2x2)N-reconstruction. \textbf{b} | Calculated electrostatic potential and DFT structure (bottom right) of the c(2x2)N-reconstruction. The blue line marks the cross section shown in Fig. 2j.}
\label{Cu100SI}
\end{figure}

\section*{Ag(111)-oxide phases}
\begin{figure}[h!]
\centering
\includegraphics[width=0.66\linewidth]{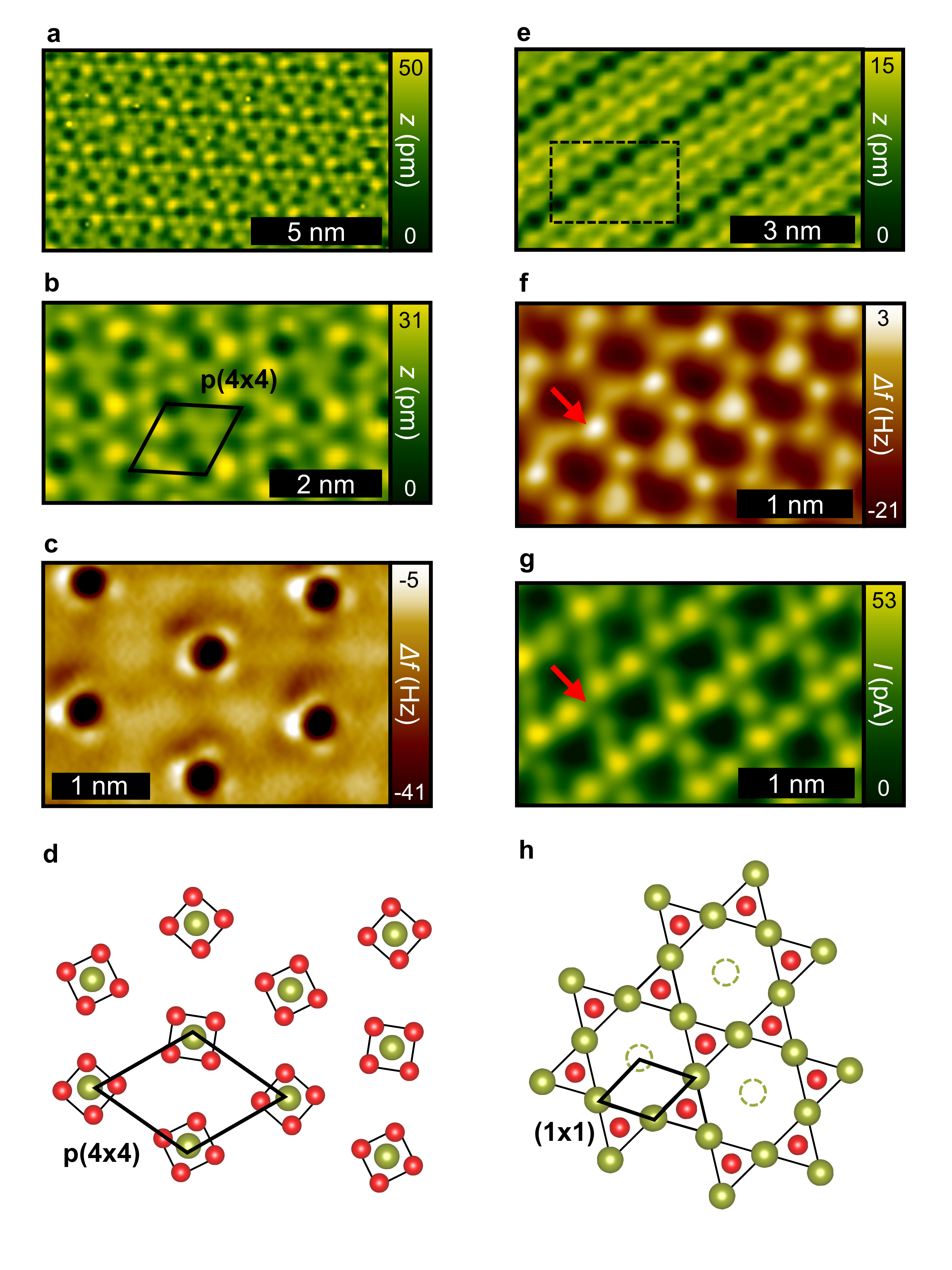}
\caption{\textbf{Ag(111)-oxide phases.} \textbf{a} | STM overview image with feedback loop (1.0 V, 50 pA) of the p(4x4)O-reconstruction on Ag(111). \textbf{b} | Small scale STM image showing the p(4x4) periodicity. \textbf{c} | CuOx-tip constant height nc-AFM measurement of the p(4x4)-reconstruction. \textbf{d} | Structural model of the p(4x4)-reconstruction derived from the nc-AFM measurement. \textbf{e} | STM overview image with feedback loop (2.0 V, 10 pA) of a stripe phase on Ag(111). The dashed rectangle marks the constant height measurement. \textbf{f} | CuOx-tip constant height nc-AFM measurement of the stripe phase. \textbf{g} | Simultaneously recorded tunneling current. The red arrow indicates the position of an oxygen atom. \textbf{d} | Structural model of the stripe phase derived from the nc-AFM and STM measurement. Ag vacancies in the pores are indicated by dashed circles.}
\label{AgOSI}
\end{figure}

Another metal-oxide system we investigated is the p(4x4)O-reconstruction on Ag(111) where several contradicting models based on combined STM/DFT data exist.\cite{Andry2024SI,AgOSI} The most recent work by Andryushechkin et al. \cite{Andry2024SI} for example proposes a structure based on Ag$_6$O and Ag$_3$O$_x$ units to explain the observed STM contrast, which shows a pronounced bias dependence.
For the preparation, a significantly higher oxygen pressure (1 x 10$^{\rm{-2}}$ mbar O$_{\rm{2}}$ for 20 min at 500 K) is required compared to the systems shown in the main paper. STM overview images of this phase are depicted in Fig. S\ref{AgOSI}a and b, while a constant-height nc-AFM measurement conducted with a CuOx-tip is displayed in Fig. S\ref{AgOSI}c. The nc-AFM data reveal the presence of elevated Ag atoms, which can be identified by the typical strongly attractive tip-sample interaction (dark spots). These dark spots are surrounded by four weak protrusions, which suggests that the basis of this reconstruction consists of AgO$_4$ units with different O-orientations with respect to the Ag(111) lattice. A schematic model of this atomic arrangement as derived from the nc-AFM measurement is shown in Fig. S\ref{AgOSI}d. 

Coexisting with the p(4x4)-phase, another silver-oxide reconstruction is observed. The overview STM image shown in Fig. S\ref{AgOSI}e reveals a linear phase modulated with dark stripes. Schnadt et al. associate this striped phase with elevated silver atoms.\cite{AgOSI} Contrary, our CuOx-tip nc-AFM data do not show distinct contrast features, which indicate the presence of any metal atoms within the top atomic layer. Yet, the nc-AFM measurement shown in Fig. S\ref{AgOSI}f reveals a surface structure which is purely terminated by hexagonal arranged protruding oxygen atoms. In addition, the simultaneously recorded tunneling current in Fig. S\ref{AgOSI}g shows Ag-vacancies in the underlying Ag(111)-lattice. By that, the structural model in Fig. S\ref{AgOSI}h can be derived, where the oxygen atoms are located on the hollow sites of the Ag(111) (1x1)-lattice. Please note the slight variations in the brightness of the oxygen atoms (Fig. S\ref{AgOSI}f), which indicate a modulation in height. This modulation is probably the origin for the striped contrast found in the STM data and could be a consequence of strain effects within the top atomic layers.

\clearpage

\section*{$\text{Fe}_{\mathbf{3}}\text{O}_{\mathbf{4}}$(110)} 
\begin{figure}[h!b]
\centering
\includegraphics[width=0.64\linewidth]{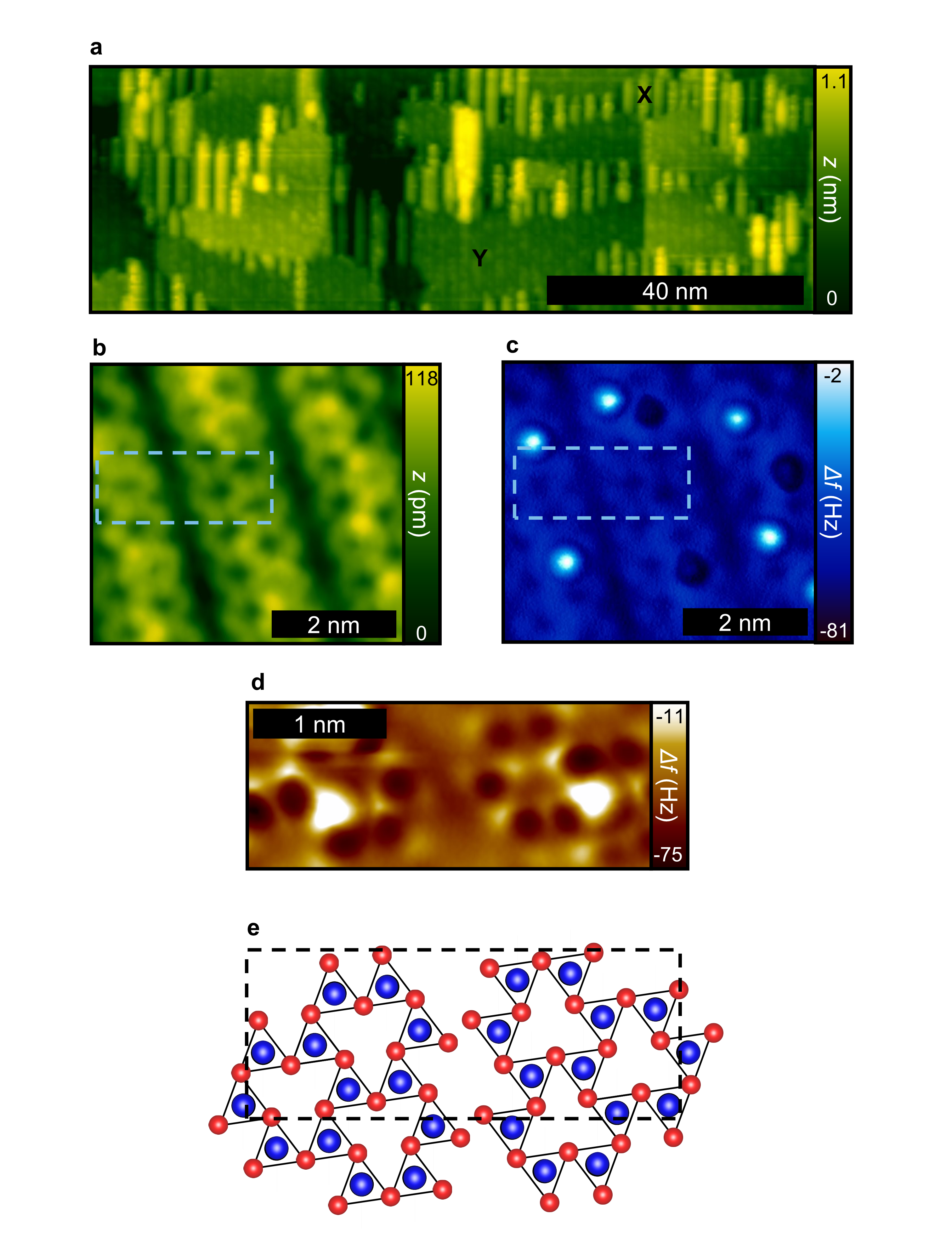}
\caption{\textbf{Magnetite $\textbf{Fe}_{3}\textbf{O}_{4}$(110).} \textbf{a} | STM overview image with feedback loop (1.5 V, 10 pA) of the $\text{Fe}_{3}\text{O}_{4}$(110) surface. Two different structures (X-phase and Y-phase) can be observed. \textbf{b} | Small scale STM image of the Y-phase. \textbf{c} | Simultaneously recorded nc-AFM with STM feedback loop. The measurement allows imaging the surface species within the pores, which show tip-induced mobility in the constant-height nc-AFM measurement and appear invisible in constant-current STM. \textbf{d} | CuOx-tip constant height nc-AFM measurement of the area in the dashed line in b and c. \textbf{e} | Structural model derived from the nc-AFM measurement of one of the strands shown in d.}
\label{MagSI}
\end{figure}

The magnetite $\text{Fe}_{3}\text{O}_{4}$(110) surface, usually features the well-known (1x3)-reconstruction and related (111) nanofacets.\cite{ParkinsonSI,JansenSI,Walls2016SI} However, as noted by two different studies,\cite{JansenSI, Walls2016SI} also a less corrugated structure forms for higher annealing temperatures (here we used 1200 K), coexisting with the faceted (1x3)-reconstruction. The overview STM image in Fig. S\ref{MagSI}a shows both phases, where the distance between the strongly corrugated stripes of the (1x3)-reconstruction (X-phase) agrees with the one between the nano-facets observed in literature.\cite{ParkinsonSI,JansenSI,Walls2016SI} The same holds for the apparent height difference of about 3 {\AA} between (1x3)-reconstructed areas and the less corrugated structure (Y-phase).\cite{Walls2016SI} A high-resolution STM image recorded on one of the flat areas (Y-phase) is shown in Fig. S\ref{MagSI}b featuring parallel strands along the [-110]-direction with regularly arranged pores. Due to mobile species on this surface, predominantly physisorbed within these pores (see also Fig. S\ref{MagSI}c), imaging in constant-height mode was challenging. Nevertheless, a small area (blue rectangle in Fig. S\ref{MagSI}b and c) could be imaged with a CuOx-tip by nc-AFM (Fig. S\ref{MagSI}d), revealing that the strands consist of threefold coordinated iron atoms. A structural model derived from the nc-AFM measurement is shown in Fig. S\ref{MagSI}e.

\clearpage

\begin{figure}[h!]
\centering
\includegraphics[width=\linewidth]{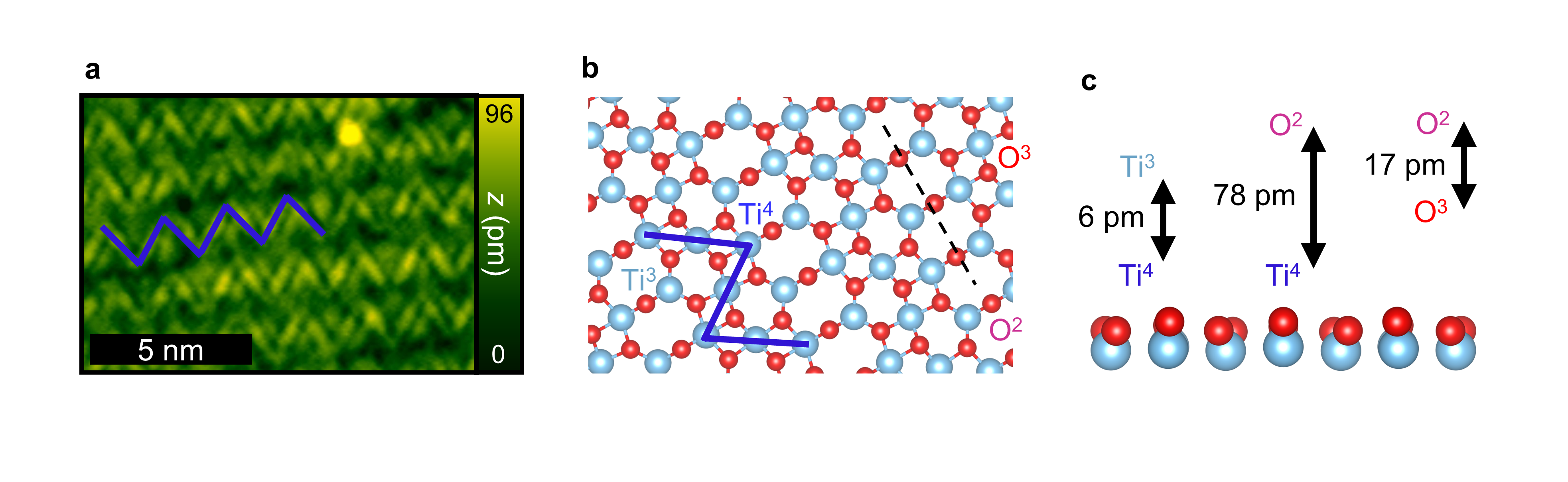}
\caption{\textbf{z-TiO$_{\rm{x}}$ on Pt(111).} \textbf{a} | STM overview image with feedback loop (2.0 V, 10 pA) showing the zigzag pattern. \textbf{b} | DFT optimized model by Barcaro et al.\protect\cite{BarcaroSI} of a similar structure to the measurement in Fig. 4 but with a smaller unit cell, i.e. with zigzag line segments with a length of 3 foufold-coordinated Ti$^{\rm{4}}$ atoms as top view. Furthermore the threefold coordinated Ti$^{\rm{3}}$ atoms are labeled as well as the threefold coordinated O$^{\rm{3}}$ atoms and the twofold coordinated O$^{\rm{2}}$ atoms. \textbf{c} Side view along the dashed line in b, depicting the relative heights of the surface species.}
\label{TiOSI}
\end{figure}

\end{document}